\documentclass[letterpaper,12pt]{article}
\usepackage[margin=1in]{geometry}

\usepackage{graphicx}
\usepackage{xcolor}
\usepackage{amsmath}
\usepackage{amssymb}
\usepackage[most]{tcolorbox}
\usepackage{subcaption}
\usepackage{hyperref}
\usepackage{multirow}
\usepackage{booktabs}

\usepackage[backend=biber,style=numeric-comp,sorting=none]{biblatex}  
\definecolor{darkgreen}{rgb}{0.0, 0.7, 0.0}

\title{\LARGE \bf
Physical Human-Robot Interaction: \\ A Critical Review of Safety Constraints }
 
\author{
Riccardo Zanella,\ \ Federico Califano,\ \ Stefano Stramigioli 
} 

\date{} 

\begin{document}

\maketitle

\footnotetext{
The authors are with the Robotics \& Mechatronics (RaM) Group, University of Twente, The Netherlands. \textit{Corresponding author: Riccardo Zanella ({\tt  schol@r-zanella.eu, r.zanella@utwente.nl}).}
}
\footnotetext{
This research has received funding from the European Union’s Horizon Europe Framework Programme under grant agreement No 101070596 (euROBIN).}


\begin{abstract}
This paper aims to provide a clear and rigorous understanding of commonly recognized safety constraints in physical human-robot interaction, particularly regarding ISO/TS 15066. We investigate the derivation of these constraints, critically examine the underlying assumptions, and evaluate their practical implications for system-level safety and performance in industrially relevant scenarios. Key design parameters within safety-critical control architectures are identified, and numerical examples are provided to quantify performance degradation arising from typical approximations and design decisions in manufacturing environments. Within this analysis, the fundamental role of energy in safety assessment is emphasized, providing focused insights into energy-based safety methodologies for collaborative industrial robot systems.  
\end{abstract}


\section{Introduction}

\begin{figure}[t]
    \centering
    \includegraphics[width=0.85\linewidth]{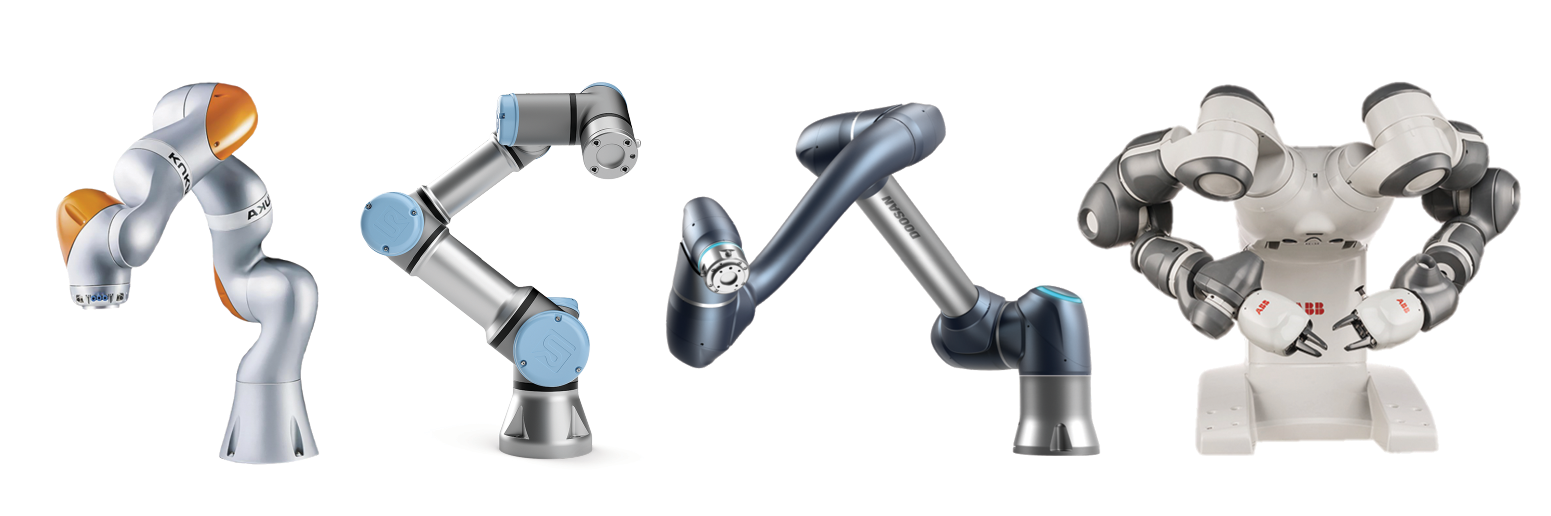}
    \caption{Examples of collaborative robots. From left to right: KUKA LBR iiwa, UR5, Doosan H2017, and ABB YuMi (courtesy of KUKA, Universal Robots, Doosan Robotics, and ABB).}
    \label{fig:cobots}
\end{figure}
Over the past two decades, collaborative robots (cobots) have emerged as a new class of robots designed to work safely alongside humans, often allowing direct physical interaction. By enabling efficient small-batch production and minimizing setup times, these systems lower the barriers to automation and thereby make flexible manufacturing both economically and technically viable, even in low-volume, high-mix production contexts. This is particularly relevant for small and medium-sized enterprises (SMEs), which have historically lacked the financial and technical resources required to deploy large-scale, dedicated robotic systems.
Operational studies indicate that human-robot collaboration can reduce cycle times by up to 50\% compared to all-human operations, while idle time decreases by 85\% when operators work alongside human-aware robots versus fully human teams \cite{shah2011fluid}.
Unlike traditional industrial robots, which are typically confined to cages or safety enclosures, cobots are engineered for open environments where intentional and unintentional contact with human operators may occur \cite{ajoudani2018progress, bi2021safety}. Their designs typically incorporate lightweight structures, rounded edges, and advanced sensor suites to support safety-focused control strategies \cite{haddadin2016physical}; representative examples of cobots are reported in \autoref{fig:cobots}.
This paradigm shift has been accompanied by growing interest in the broader coexistence of humans and robots and in their mutual exchange of information, encompassed by the notion of Human-Robot Interaction (HRI). HRI includes all forms of mutual influence, from non-collaborative scenarios such as supervision, teleoperation, or social interaction, to direct collaboration. Within this spectrum, a further distinction arises when the interaction involves physical contact, leading to the notion of \emph{physical} Human-Robot Interaction (pHRI), where the emphasis is on the physical component of the interaction. Accordingly, the forces, energy, and power exchanged at the physical interface are key quantitative metrics for safety assessment, regulatory compliance, and control design. It is worth noting that cases in which the physical component is completely absent are relatively rare: even with careful trajectory planning and contact-avoidance algorithms, uncertainty and unforeseen human motion imply a residual possibility of contact. 

A primary concern in scenarios involving pHRI is ensuring human safety, even under dynamic and partially unknown environmental conditions. 
While \textit{safety} is often treated as a fundamental property, it is an emergent characteristic of the interaction context: it depends not only on the robot's behavior but also on the environment, the task, and human actions. For instance, should safety be operationally defined as the prevention of physical injury, or more broadly as the avoidance of pain and discomfort? Furthermore, can a system legitimately be classified as safe if there exists a small but nonzero probability of harm? In some cases, defining or assessing safety becomes a philosophical problem. For example, what can a robot do to remain “safe” if a human suddenly runs toward it and collides, even when the robot itself is stationary and behaving correctly? 

To provide practical guidance, the International Organization for Standardization (ISO) helps to bridge the gap between the abstract concept of safety and its operational implementation by adopting a functional perspective in which safety is regarded as the capability of a robotic system to operate within limits that ensure risks to humans are reduced to an acceptable level. These limits are typically expressed in terms of forces, velocities, accelerations, energy transfer, or spatial constraints, thereby providing quantifiable and enforceable metrics for safe operation.
In 2011, the ISO released the standard ISO 10218-2 \cite{standardISO10218_2011}, which established foundational safety requirements for the integration of industrial robots into human work environments. Building on this, the technical specification ISO/TS 15066 \cite{standardISO15066} was introduced in 2016 to support practitioners in ensuring safety in HRI. 
A concrete risk comes from the belief that cobots’ design makes them inherently safe, resulting in inadequate control measures, especially in dynamic and complex industrial settings. To this purpose, in 2025, ISO 10218-2 was revised to deleverage on the concept of cobots in general: only the application can be designed, verified, and validated as collaborative, rather than the robot \cite{standardISO10218}. Besides this conceptual shift, the new standard confirms the core methods for quasi-static and transient contact in Power and Force Limiting mode originally defined in ISO/TS 15066 by incorporating them unchanged into Annex M of ISO 10218-2:2025. As a result, both documents are technically equivalent regarding these safety limits and may be used interchangeably for the purposes of this work. Nevertheless, we continue to cite ISO/TS 15066, as it remains the predominant reference for these technical guidelines in pHRI and future revision will preserve the same identifier. The technical specification ISO/TS 15066 provides quantitative limits to mitigate the risk of human injury during collaborative operations. It enables the dynamic coexistence of humans and robots in the same workspace by defining the allowable motion characteristics under four different collaborative operation modes. Among these modes, the only one that accounts for physical interaction between the robot and a human is the \textit{Power and Force Limiting} (PFL) mode, which presents a pre-collision guideline aimed at producing safe pHRI. 

Despite the concrete recommendations provided by these standards, they contain inherent ambiguities that may lead to inconsistent risk assessments and potential underestimation of hazards \cite{kirschner2022iso}. This highlights a broader concern: while the standards offer quantifiable limits and operational definitions, safety in pHRI remains context-dependent, emergent, and sensitive to specific tasks and environmental conditions. ISO/TS 15066 provides a baseline framework whose application requires careful interpretation, as assumptions embedded in the standard may affect both safety assessments and system performance \cite{kirschner2021notion}, emphasizing the need for context-aware evaluation.  A proper understanding is crucial not only to implement safety coherently and minimize risks but also to optimize robot performance. 
In industrial applications, the objective is to ensure that robotic systems perform their designated tasks with maximal operational efficiency while rigorously preserving human safety. Effective safety strategies, therefore, demand a careful balance between risk mitigation and operational efficiency: a robot should not only avoid unsafe behavior (which, according to the standards, could be trivially achieved by keeping it switched off) but must also perform its assigned tasks effectively. While this consideration applies broadly to all practitioners, it is particularly essential for researchers developing novel control strategies, safety frameworks, or pHRI paradigms, where a deep understanding of task-specific dynamics, robot design, and human variability is required to ensure both safe and efficient operation.

In this perspective, the present paper aims to provide a clearer understanding of commonly recognized safety constraints in pHRI. This objective is pursued by presenting a clear, self-contained, and didactic reconstruction of the derivation process, thereby offering a coherent and unified view of the concepts and assumptions underlying their formulation, which are often presented in a fragmented manner in the literature. Such clarification is essential for correctly interpreting the standards, applying them consistently, and critically assessing their implications. It also enables a more transparent analysis of the trade-off between safety guarantees and robotic performance, ultimately supporting the identification of more informed design choices and best practices. 

A substantial body of literature has reviewed safety in HRI and pHRI from complementary perspectives, reflecting the multidisciplinary nature of the problem. Early works focused on organizing interactions according to contact type and task context, highlighting implications for both control design and robot safety \cite{de2008atlas}, discussing control strategies for pHRI, and considering mechatronic design aspects of collaborative robots \cite{haddadin2016physical}. These contributions established foundational links between robot mechanics, contact forces, and human safety; however, they were published before the widespread  adoption of collaborative robots and the consolidation of ISO/TS 15066 as a reference framework.
More recent surveys have systematically organized safety strategies according to perception, planning, and control layers. They mainly discuss sensing technologies, human intention estimation, and safety-oriented motion planning, often framing safety as a combination of collision avoidance, speed and separation monitoring, collision detection, and post-impact mitigation \cite{li2024safe, alenjareghi2024safe}. While these reviews provide valuable methodological overviews, they still lack in-depth analysis of safety constraints themselves, often treating them as externally given inputs rather than as objects of analysis. 
Other reviews explicitly focus on standards and regulations in collaborative robotics, discussing ISO 10218 and ISO/TS 15066, their scope, and limitations \cite{martinetti2021redefining, li2024safe}. These works highlight ambiguities in the interpretation of biomechanical limits, variability across human subjects, and challenges in translating tabulated thresholds into real-time control constraints. However, they typically do not reconstruct how these limits are derived or critically assess the underlying modeling assumptions, nor do they systematically analyze the safety-performance trade-off.
This work addresses this gap by placing the derivation and interpretation of safety constraints at the center of the analysis, with explicit attention to how modeling choices affect both safety margins and robotic performance in pHRI.

Building on the importance of understanding safety constraints in context, it is useful to consider energy as a unifying metric for pHRI safety. While much of the literature and industrial practice in HRI safety still relies on the traditional paradigm of physical separation between humans and robots, an effective strategy for pHRI safety focuses on monitoring and controlling the physical energy exchanged between the robot and its environment \cite{haddadin2012truly,haddadin2008evaluation}. Energy is a fundamental concept in physics, providing intuitive insight and practical tools that have driven scientific and engineering advances \cite{ortega2002putting}. In robotics, physical energy flows encode information on performance, safety, and efficiency, making energy a key factor for reconciling task-based optimization with meaningful safety guaranties in dynamic human-robot interactions.
Strategies aimed only at avoiding contact are inherently limited in pHRI scenarios, where physical interaction may be intentional, unavoidable, or even required for task execution. During such interactions, both information and energy are transmitted bidirectionally through the robot-environment coupling, making energy a natural and physically interpretable measure of potential risk. Framing safety in energetic terms allows robotic systems to continuously monitor and modulate their energy flows in real time, providing a metric that is both interpretable and directly linked to potential human harm.
Importantly, this energetic perspective is explicitly supported by ISO/TS 15066 in the PFL mode, which establishes a quantitative relationship between biomechanical pain or injury thresholds and energy transfer during transient contact. By connecting mechanical energy to human injury metrics, this approach provides a rigorous and physically interpretable foundation for designing control strategies that are both safe and effective in pHRI scenarios. 
Based on this understanding, the present work aims to examine the role of energy in safety assessment and the design of energy-based control strategies, offering targeted insights on existing literature, clarifying common misconceptions, and outlining promising directions for future research. This highlights how energetic approaches can link theoretical safety guaranties with practical, high-performance robotic operation. 
\\\\
\noindent
Motivated by the above considerations, this work makes the following contributions:
\begin{itemize}
    \item A systematic derivation of safety constraints for pHRI within the commonly accepted regulatory framework (i.e., the PFL mode specified in ISO/TS 15066).
    \item A clear identification and critical justification of the assumptions underlying the constraints, together with an in-depth analysis of the entire derivation process.
    \item An analysis of the implications of these constraints on both human safety and robotic efficiency in pHRI, investigating the interdependence among the different parameters and discussing their impact on safety margins and task performance. Based on recent research evidence and dedicated considerations, potential margins for improvement are also identified.
    \item A concise review of the role of energy in safety assessment and in the design of energy-based control strategies, including targeted remarks on the existing literature, clarification of common misconceptions, and an outline of promising directions for future research. 
    \item  Identification of key design choices that directly affect safety and performance in pHRI, accompanied by a quantitative investigation of their effects on safety enforcement. 

\end{itemize}

The remainder of the document is structured as follows. In \autoref{sec:robot_constraints}, we present the rationale and steps for deriving operational safety constraints on robot motion based on empirical  measures of human pain sensitivity. In \autoref{sec:discussion}, we discuss the resulting constraints and their derivation process, highlighting key considerations for designing practical safety protocols, potential limitations that may affect safety and performance, and the interdependence of parameters. Finally, in \autoref{sec:energy}, we examine the role of energy more broadly, situating it within the literature on risk assessment and methodological implementation, and emphasizing conceptual and design aspects of energy-based approaches.

 
\section{Operational Constraints of Robotic Motion }
\label{sec:robot_constraints}

Ensuring safety in pHRI fundamentally requires the quantitative characterization of human pain and injury thresholds, as well as the systematic mapping of these limits onto the corresponding robot-side variables that must be constrained during contact. The underlying principle is conceptually simple: once the maximum tolerable biomechanical load for a specific anatomical region has been determined, the robot must be controlled such that no interaction surpasses this threshold. ISO/TS 15066 formalizes this principle by representing human safety in terms of maximum allowable force and pressure values for each body region. These threshold values are derived from two research streams. The FP-0317 research project \cite{FP0317} at the University of Mainz quantified quasi-static pain-onset thresholds in $100$ healthy adult participants across $29$ anatomical regions using a standardized algometry protocol, thereby establishing body-region-specific force and pressure limits. In parallel, the DGUV/BGIA recommendations \cite{BG_BGIA_2009} provided earlier methodological frameworks, reference measurements, and broader criteria for acceptable force levels associated with a low risk of minor injury. 
These studies consider only quasi-static conditions, characterized by slow and sustained contact. However, a physical robot–human collision typically occurs in two phases: an initial transient contact, corresponding to a brief impact governed by the robot- and human-reflected dynamics, followed by a quasi-static contact.  Consequently, ISO/TS 15066 extends the quasi-static injury thresholds to transient, impulsive contacts by introducing a scaling factor that accounts for the increased tolerance of human tissues to short-duration loading. In the standard, transient contacts are defined as lasting less than $0.5,\mathrm{s}$, for which the allowable force and pressure limits are set to at least twice the corresponding quasi-static values, consistent with experimental and biomechanical evidence \cite{mewes2003safeguarding, suita1995failure}.

As previously noted, in order to render these tolerance values operational, they must be reformulated as explicit constraints on the robot’s motion. In ISO/TS 15066, human tissue is modeled as a linear elastic spring, enabling the derivation of a direct relationship between pain sensitive thresholds (expressed in terms of force or pressure) and the maximum elastic potential energy that the affected body region can safely store during deformation. This energy bound can subsequently be mapped to a corresponding upper limit on robot speed. The remainder of this section presents a formal and detailed derivation of the proposed mapping.

\begin{figure}[t]
    \centering
    \begin{subfigure}[b]{0.3\linewidth}
        \centering
        \includegraphics[width=0.85\linewidth]{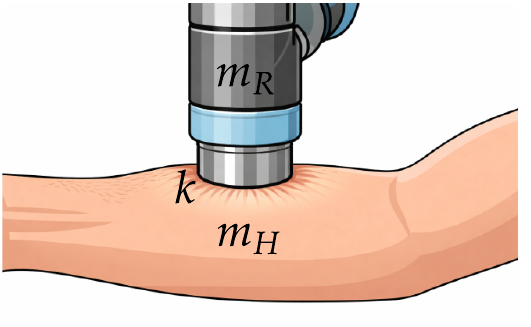}
        \includegraphics[width=\linewidth]{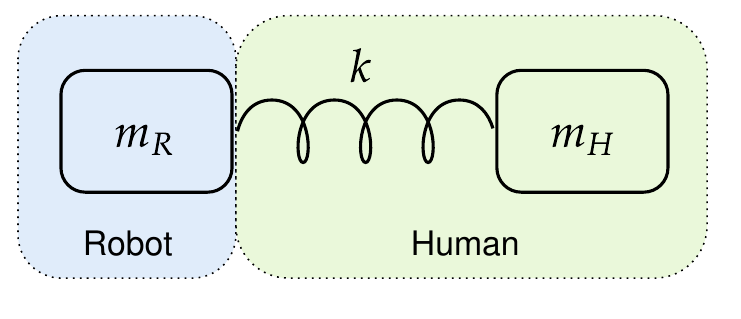}
        \caption{}
        \label{fig:model_mass_spring_mass}
    \end{subfigure} 
    \begin{subfigure}[b]{0.5\linewidth}
        \centering
        \includegraphics[width=\linewidth]{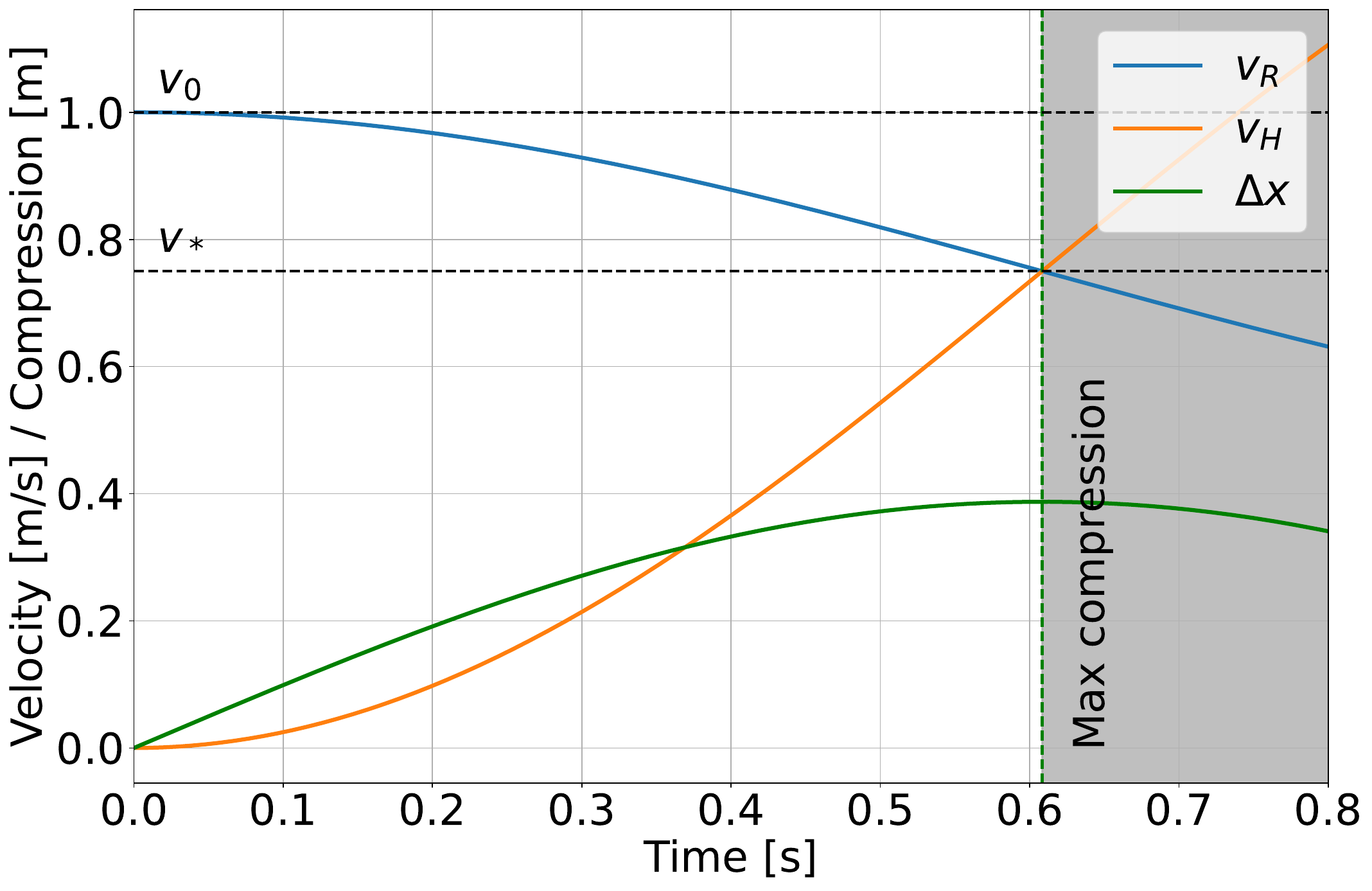}
        \caption{}
        \label{fig:plot_mass_spring_mass}
    \end{subfigure} 
    \caption{(a) Mechanical model representing a robot-human collision: the moving robot body is modeled as an effective mass $m_R$, the impacted human body region as an effective mass $m_H$, and tissue deformation at the contact point is represented by a spring of stiffness $k$. (b) Simulation results with $m_R = 3.0$~kg, $m_H = 1.0$~kg, $k = 5.0$~N/m, and initial velocity $v_0 = 1.0$~m/s. The plot displays the time evolution of the velocities of the masses $m_R$ and $m_H$, denoted $v_R$ and $v_H$, respectively, as well as the spring compression $\Delta x$. The maximum spring compression occurs after $0.6~\mathrm{s}$, at which point both masses attain approximately the same velocity, $v_* \approx 0.75~\mathrm{m/s}$. The shaded area indicates motion after maximum compression.}
    \label{fig:mass_spring_mass_overall}
\end{figure}

A fundamental preliminary step consists of establishing a model for human-robot contact. In accordance with ISO/TS 15066 and a substantial portion of the relevant literature, the interaction is typically idealized as a collision between two lumped masses interconnected by a compliant element, as depicted in \autoref{fig:model_mass_spring_mass}. Note that this one-dimensional collision model corresponds to a mapping of the interaction between a multi-articulated robotic system and a spatially distributed human body onto the direction of contact. This mapping leads to a scalar formulation in which the interaction dynamics is governed by equivalent masses along the contact direction, representing the portions of the system that effectively participate in the interaction. Specifically, the robot is reduced to an effective mass $m_R$, which denotes the configuration- and direction-dependent reflected inertia of the manipulator at the contact point. Analogously, the impacted region of the human body is modeled by an effective mass $m_H$, which characterizes the apparent inertia of the corresponding anatomical segment involved in the interaction. Within the same formulation, the deformation of biological tissues at the contact interface is modeled by a linear elastic spring with stiffness $k$, which accounts for the elastic energy stored during the impact event

In this setting, the robot $m_R$ approaches the human $m_H$ with an initial velocity $v_0$, while the human is assumed, without loss of generality, to be initially at rest, i.e., $v_H=0$; if the human is moving, $v_0$ should be interpreted as the relative velocity of the rigid body with respect to the human at the moment of impact.
In the absence of external forces acting on the system, the total linear momentum remains conserved. Denoting by $v_*$ the common velocity of the bodies at the instant of maximum compression, the principle of conservation of linear momentum can be expressed as:
\begin{equation*}
    \begin{aligned}
    m_R v_0 + m_H \cdot 0 &= (m_R + m_H) v_* \\
    &\Rightarrow v_* = \frac{m_R}{m_R + m_H} v_0.
    \end{aligned}
\end{equation*}
The initial kinetic energy of the system is entirely due to the moving robot mass:  
\begin{equation*}
    K_0 = \frac{1}{2} m_R v_0^2.
\end{equation*} 
As the robot comes into contact, the spring compresses until the two masses reach a state where their velocities coincide. At the instant of maximum spring compression, the relative velocity between the two bodies vanishes, and both move together with the same velocity, as shown in \autoref{fig:plot_mass_spring_mass}. 
Hence, the kinetic energy at maximum compression is   
\begin{align*}
    K_* &= \frac{1}{2} (m_R + m_H) v_*^2  = \frac{1}{2} (m_R + m_H) \left(\frac{m_R}{m_R + m_H} v_0 \right)^2 = \frac{1}{2} \frac{m_R^2}{m_R + m_H} v_0^2 
\end{align*}  
At this moment, part of the robot's kinetic energy from the free motion is stored in the spring as elastic potential energy $U_s$:   
\begin{equation}
    \label{eq:energy_balance_impact}
    \Delta K =  U_s.
\end{equation}
The magnitude of this energy change, $\Delta K$, is the difference between the robot’s initial kinetic energy and that of the combined system after the impact. It can be expressed directly in terms of the masses and the robot’s initial velocity as
\begin{align*}
    \Delta K &= K_0 - K_*  =\frac{1}{2} m_R v_0^2 - \frac{1}{2} \frac{m_R^2}{m_R + m_H} v_0^2 = \frac{1}{2} \frac{m_R m_H}{m_R + m_H} v_0^2.
\end{align*}
The elastic potential energy $U_s$ stored in a linear spring depends on its compression $\Delta x$ and on the instantaneous force that the spring exerts on the tissues. According to Hooke's law, this force is 
$F_s = k \, \Delta x$, so that the energy can be expressed as
\begin{equation*}
    U_s = \frac{1}{2} k \, \Delta x^2 = \frac{F_s^2}{2k}.
\end{equation*}

From biomechanical studies, empirically determined values of the maximum force that human tissues can safely tolerate are available, denoted as $F_{\max}$, or alternatively as the corresponding pressure $p_{\max}$, which is related to the force by the contact area $A$ through $F_{\max} = A \, p_{\max}$.
Given these values, the maximum elastic potential energy that can be stored in the tissue during the collision is  
$$
    U_{s,\max} = \frac{F_{\max}^2}{2k},
$$
where $k$ is the effective stiffness of the tissue. Assuming no energy dissipation through friction or other mechanisms, this energy corresponds to the maximum kinetic energy that can be transferred from the robot to the human body during impact, i.e.,  $\Delta K_\text{max} = U_{s,\max}$, which occurs at the point of maximum tissue compression. This formulation establishes a direct relationship between the force experienced by the tissue and the maximum admissible velocity of the robot prior to impact, providing a quantitative link between pain sensitive thresholds and robot motion constraints, 
$$
\frac{1}{2} \frac{m_R m_H}{m_R + m_H} v_{0,\max}^2 = \frac{F_{\max}^2}{2k},
$$  
which leads to the maximum admissible robot velocity before impact:
\begin{equation} \label{eq:lim_speed}
    \begin{aligned}
        v_{0,\max} &= \sqrt{\frac{2 \, \Delta K_\text{max} (m_R + m_H)}{m_R m_H}}  = \sqrt{\frac{m_R + m_H}{m_R m_H}} \frac{F_{\max}}{\sqrt{k}}  = \sqrt{\frac{m_R + m_H}{m_R m_H}} U_{s,\max}.
    \end{aligned}
\end{equation}

This expression provides a direct link between pain sensitive thresholds and robot motion parameters: it specifies the maximum impact velocity that guaranties tissue forces remain below the critical level $F_{\max}$. Furthermore, knowing this velocity, the corresponding upper bound on the robot’s initial kinetic energy can be computed as  
\begin{equation*} \label{eq:lim_energy}
    \begin{aligned}
        K_{0,\max} &= \frac{1}{2} m_R v_{0,\max}^2  = \frac{m_R + m_H}{m_H}\frac{F_{\max}^2}{2k}  = \frac{m_R + m_H}{m_H} U_{s,\max}
    \end{aligned}
\end{equation*}

\begin{figure}
    \centering
    \includegraphics[width=0.7\linewidth]{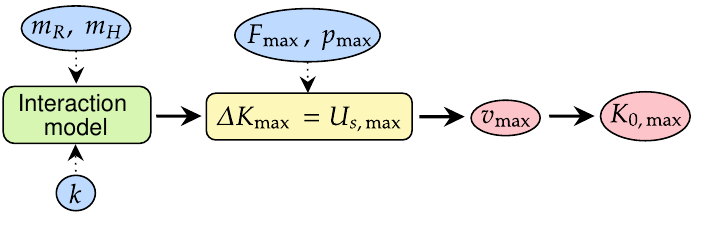}
    \caption{Schematic pipeline for deriving robot motion safety limits in terms of maximum admissible speed $v_{\max}$ and pre-collision kinetic energy $K_{0,\max}$ (red circles) from force- and pressure-based pain thresholds. Solid arrows denote the sequence of analytical derivations, while dashed arrows indicate the dependencies on parameter (blue circles): robot effective mass $m_R$, human effective mass $m_H$, body-region stiffness $k$, and maximum allowable force or pressure before pain $F_{\max}$/$p_{\max}$. The interaction is modeled using the mass-spring-mass representation shown in \autoref{fig:model_mass_spring_mass}.   }
    \label{fig:pipeline}
\end{figure}

In conclusion, from an operational standpoint, it is required that at least one of the following equivalent constraints be satisfied during the entire robot’s motion:
\begin{align}
    &\Delta K\leq U_{s,\max} \label{eq:constr_deltak}\\
    & K_0 \leq K_{0, \max}  \label{eq:constr_k}\\
    & v_{0} \leq v_{0,\max} \label{eq:constr_v}
\end{align}
The complete pipeline that maps pain sensitive thresholds to operative safety constraints on robot motion is schematically depicted in \autoref{fig:pipeline}.
As shown, this procedure is critically dependent on the assumptions underlying the interaction model, the experimentally derived force and pressure thresholds $F_{\max}, p_{\max}$ \cite{FP0317,BG_BGIA_2009}, the effective stiffness of the biological tissues $k$, and the effective masses of the robot and human body segments $m_R$ and $m_H$, respectively. In the subsequent section, we analyze in detail how these parameters and modeling assumptions influence both safety and task performance, highlighting their relationship to the technical specification and the related literature.

\section{Remarks on Deriving Robot Motion Constraints} \label{sec:discussion} 
In this section, we take a deeper dive into the derivation pipeline introduced in \autoref{sec:robot_constraints}, delivering a more granular interpretation of the procedure and its outcomes. We explicate the meaning and operational consequences of the principal simplifying assumptions, highlight how these modeling choices propagate through the system to influence both performance and safety, and identify concrete design parameters that can be tuned when implementing a safety-critical control architecture.

\subsection{Derivation process  and fully inelastic assumptions}

It is important to notice that the results detailed in \autoref{sec:robot_constraints} follow directly from the constraint formulation in ISO/TS 15066, while making explicit the intermediate steps required to derive them.
We also emphasize that the constraints derived in \autoref{sec:robot_constraints} define operative limits on the robot’s motion in terms of the maximum admissible speed (or, equivalently, kinetic energy) \emph{before impact} , such that the force or pressure generated \emph{during the collision} does not exceed the tolerable pain thresholds for the contacted human body region.
By representing the interaction as a collision between two discrete masses and explicitly accounting for the local mechanical stiffness of the impacted human body region, the critical time instant for safety assessment corresponds to the moment of \emph{maximum deformation}, at which the biomechanical loading reaches its peak value. The entire derivation pipeline, shown in \autoref{fig:pipeline}, is formulated at this instant. As a result, the only phase relevant for safety is the time interval before maximum compression. The resulting constraints do not require nor depend on any assumptions about what happens after this peak load. Unsafe modes are triggered by the maximum deformation alone; whether the collision dissipates energy (inelastic) or returns it (elastic) is irrelevant because such post-maximum-load dynamics occur only after the critical threshold has already been reached.
In contrast, ISO/TS 15066 and prior works such as \cite{rossi2015pre, meguenani2015control} derive the limits from the difference between the kinetic energy \emph{before and after impact}, explicitly assuming a \emph{fully inelastic collision} in which all lost kinetic energy is dissipated through tissue deformation or fracture. This assumption is strong and not generally representative, as real human-robot impacts occur between the extremes of perfectly elastic and perfectly inelastic behavior, raising concerns about the conservativeness of the fully inelastic model \cite{benzi2023energy}.
In summary, \textbf{the robot’s motion constraints} (identical to the one reported in ISO/TS 15066) \textbf{should be interpreted as an instantaneous safety condition}: the robot must enter the collision with sufficiently low kinetic energy to prevent the tissue from exceeding its allowable maximum compression. \textbf{The restitution characteristics of the impact, regardless of how elastic or inelastic they may be, do not alter this requirement}.

\subsection{Pain Thresholds}  \label{sec:pain_thresholds} 

In this remark, we observe from \eqref{eq:lim_speed} that the maximum admissible speed, \(v_{0,\max}\), exhibits a linear dependence on the force bound \(F_{\max}\) and an inverse dependence on the stiffness parameter \(k\). Consequently, if a collision involves a body region characterized by lower effective stiffness or higher maximum tolerable force, the robot can operate at a higher allowable velocity, since increased compliance or more relaxed force limits permit greater energy transfer without surpassing safety thresholds. 
Furthermore, it should be emphasized that the maximum admissible pressure $p_{\max}$ is typically determined from biomechanical studies, whereas the corresponding maximum allowable interaction force $F_{\max}$ is obtained according to $F_{\max} = A p_{\max}$ for a specific contact area $A$, which in turn depends on the robot’s geometry, its end-effector equipment, and its specific configuration during physical interaction. Directly using the $F_{\max}$ values reported in technical specifications may appear more convenient, as it removes the need for an explicit determination of $A$, which can indeed be difficult to assess. However, these tabulated values are typically computed under the assumption of a conservative nominal contact area of $A \leq 1 \mathrm{cm^2}$. Consequently, \textbf{when a more accurate estimate of the actual contact area is available, it is advantageous to adjust $F_{\max}$ accordingly}. 
It is important to note that the \textbf{admissible robot velocity exhibits a linear dependence on the contact area}. As a result, for instance, doubling the contact area to $2 \mathrm{cm^2}$ allows the robot to operate at twice the velocity associated with the nominal $F_{\max}$ specified for $A = 1 \mathrm{cm^2}$.
Collaborative robots are intentionally designed to minimize sharp edges and protrusions, which reduces local pressure and lowers the risk of injury. However, when additional equipment (e.g., grippers or tools) is attached to the robot, or when sharp objects are being manipulated, the effective contact area can be reduced, thereby lowering the maximum safe force that can be applied during a collision. In such situations, control strategies and trajectory planning can be used to increase safety. 

At the beginning of this section, we provided a concise overview of the origin of the maximum force and pressure limits specified in ISO/TS 15066. However, the technical specification itself acknowledges that these limit values are derived from a restricted set of experimental data and are therefore subject to revision as additional studies, specifically focused on collaborative robotics, become available. Rather than attempting an exhaustive review, we underscore this point by referencing some representative works that contribute pertinent new empirical evidence. 
In particular, these studies: (i) report controlled impact experiments that directly quantify force thresholds under dynamic collision conditions \cite{han2024evaluation}; (ii) differentiate between pain onset and maximum tolerable pain, providing statistically characterized thresholds across multiple anatomical regions instead of single deterministic limits \cite{han2022assessment}; and (iii) examine pressure-based metrics under localized contact, demonstrating how contact area and spatial pressure distribution modulate pain perception during robot-human collisions \cite{park2019assessment}. 
Collectively, these and related studies advance the state of the art beyond ISO/TS 15066 by replacing quasi-static, worst-case tabulated limits with experimentally derived, body-region-specific, statistically supported pain thresholds. These thresholds match dynamic safety assessment approaches more closely and can be more directly embedded into control-oriented safety frameworks for collaborative robots.

\subsection{Human Effective Mass} \label{sec:human_effective_mass}

The key dependence on the robot and human masses enters through the reduced-mass combination $\frac{m_R + m_H}{m_R m_H} = \frac{1}{m_R} + \frac{1}{m_H}$.
This expression shows that the maximum allowable impact velocity $v_{0,\max}$ is limited by the combined inertia of the robot and the human body segment along the contact direction. For a given human mass $m_H$ and tissue stiffness $k$, decreasing the effective robot mass $m_R$ reduces the reduced-mass term, directly increasing the permissible velocity. Lighter robot components or configurations with lower reflected inertia along likely contact directions therefore allow higher safe impact speeds, while heavier robot masses increase the reduced mass, lowering $v_{0,\max}$.
This insight provides clear guidance for safety-oriented design and control. Collaborative robots are typically designed with lightweight links and low-inertia structures to exploit this effect, improving safety while preserving performance. For redundant manipulators, additional safety can be achieved through control strategies that dynamically adjust the robot’s configuration to reduce its reflected mass in specific directions \cite{rossi2015pre, steinecker2022mean}. By selecting joint configurations or motion strategies that minimize the inertia along a predicted collision path, the robot can safely move faster while keeping the energy transferred during a collision within pain sensitive thresholds. In summary, both mechanical design and active control of reflected inertia are efficient levers to maximize allowable velocities without exceeding tissue force thresholds.

In order to understand the influence of the robot and human masses on the maximum allowable impact velocity, we examine extreme cases of the reduced-mass combination. When the robot mass dominates ($m_R \gg m_H$), the reduced-mass factor simplifies to  
\begin{align*}
\frac{m_R+m_H}{m_R m_H} \approx \frac{1}{m_H} 
\quad\Rightarrow\quad \begin{matrix}
v_{0,\max} \approx \frac{F_{\max}}{\sqrt{k\,m_H}},\\ 
K_{0,\max} \approx \frac{m_R}{m_H} U_{s,\max}.
\end{matrix} 
\end{align*}
In this case, the admissible impact velocity depends primarily on the human effective mass and tissue stiffness, while the initial robot kinetic energy can be much larger than the tissue energy capacity. The residual robot energy not absorbed by the tissue can produce secondary motion or post-impact accelerations, emphasizing the importance of considering both peak force and post-impact dynamics in safety-oriented design.

\begin{figure}
    \centering
    \includegraphics[width=0.99\linewidth]{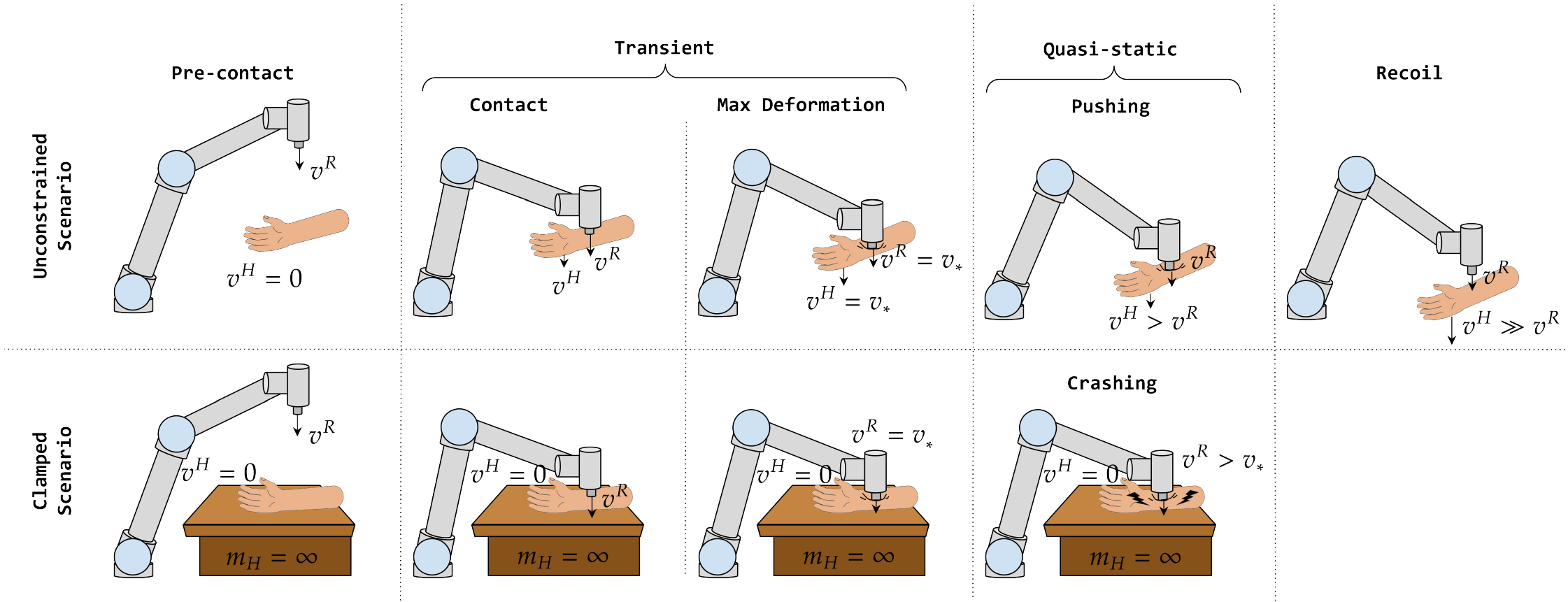}
    \caption{A physical robot–human collision typically occurs in two phases: a transient contact (brief impact governed by robot- and human-reflected dynamics) followed by a quasi-static contact, where the contact can be further subdivided into either pushing (unconstrained) or crushing (clamped) \cite{haddadin2016physical}. The figure illustrates unconstrained and clamped interaction scenarios. In both cases, the robot approaches with initial velocity $v^R>0$ while the human is initially at rest ($v^H=0$). In the unconstrained case, the human accelerates after impact and eventually recoils, resulting in a pushing interaction with decreasing relative velocity. In the clamped case, the human remains stationary throughout, leading to a crushing interaction.}
    \label{fig:contact_type}
\end{figure}

Conversely, when the human effective mass is much larger than the robot mass ($m_H \gg m_R$), the reduced-mass factor becomes  
\begin{equation} \label{eq:clamped_scenario}
    \begin{aligned}
     \frac{m_R+m_H}{m_R m_H} \approx \frac{1}{m_R} 
    \quad\Rightarrow\quad \begin{matrix}
    v_{0,\max} \approx \frac{F_{\max}}{\sqrt{k\,m_R}}, \\
    K_{0,\max} \approx U_{s,\max}.
    \end{matrix}
    \end{aligned}
\end{equation}
The limiting case, i.e., for $m_H =\infty$, can be interpreted as a \textit{clamped condition} in which the human body segment is constrained and therefore unable to recoil in response to the impact, as depicted in \autoref{fig:contact_type}. In this situation, nearly all of the kinetic energy that is transferred from the robot to the human during the collision is stored as elastic energy in the tissues, with no motion of the human segment. The corresponding maximum allowable initial kinetic energy of the robot is therefore approximately equal to the tissue energy capacity.
This scenario is thus more conservative than the general case in which the human operator is able to recoil: for the same robot mass and tissue stiffness, allowing the human segment to move ($m_H$ finite) increases the reduced-mass term in the velocity formula, which in turn increases $v_{0,\max}$. Consequently, designing the control to respect safety speed (or energy) constraints under the potential-clamp assumption ensures that peak forces at the tissue do not exceed pain-sensitive thresholds under any realistic human mobility condition, providing a robust safety margin.

It should be noted that ISO/TS 15066 does not explicitly consider the clamped case in the constraints. Instead, the standard distinguishes only between quasi-static and transient contacts by applying empirical multiplicative factors to the admissible force limits, without modifying the underlying inertial interaction model. 
However, based on the preceding discussion, \textbf{it is more appropriate to employ \eqref{eq:clamped_scenario} for conditions in which the operator’s body segment is effectively immobilized, and to use \eqref{eq:constr_k} or \eqref{eq:constr_v} for more general contact interactions in which the operator is able to recoil or withdraw from the moving robot}.

In the general case, where $m_R$ and $m_H$ are comparable, the maximum allowable robot velocity is given by \eqref{eq:lim_speed} and is bounded by
\begin{equation}
    \frac{F_{\max}}{\sqrt{k\, \max(m_R, m_H)}} < v_{0,\max} < \frac{F_{\max}}{\sqrt{k\, \min(m_R, m_H)}}.
\end{equation}
This shows that the maximum velocity increases when the lighter body is the robot, allowing for faster motion without exceeding the tissue force limit, and decreases when the lighter body is the human, reflecting the reduced capacity of the human segment to absorb energy. The intermediate case highlights the joint influence of robot and human effective masses on safe impact velocities and provides a practical guideline for reducing robot inertia along likely contact directions to maximize allowable speed while maintaining biomechanical safety.
  
The preceding analysis demonstrates that obtaining an accurate estimate of the human effective mass is essential for conducting reliable and effective safety assessments. The technical specification provides specific values that denote biomechanical equivalent masses derived from dynamic impact responses and scaled from anthropometric reference data originating from \cite{USAF1988}. The specification adjusts these values to reflect the portion of mass that effectively participates in momentum exchange during a collision (dependent on posture, joint coupling, and available recoil) while incorporating conservative safety margins.
For example, although the anatomical mass of the forearm is approximately $1.4\,\mathrm{kg}$ for a 50$^{\text{th}}$-percentile male \cite{USAF1988}, ISO/TS~15066 assigns it an effective mass of $2\,\mathrm{kg}$, since part of the upper arm and shoulder contributes to dynamic recoil. Similarly, the anatomical mass of the thigh (about $9.8\,\mathrm{kg}$) is replaced by an effective mass of $75\,\mathrm{kg}$. This is because a lower-limb impact often involves nearly the entire body mass, and the standard specifies that \textit{``Mass values for thighs, knees, and lower legs are set to the full body weight, since these body parts are not free to recoil or retract from impact while the operator is standing.''}

\subsection{Robot Effective Mass} \label{sec:robot_effective_mass}
As already shown before in the document, the robot’s effective mass $m_R$ plays a central role in determining the severity of human injury in the event of a collision, as it directly influences the kinetic energy transferred to the human body. Accurate estimation of this mass is therefore critical for both the design and control of safe pHRI. However, evaluating the instantaneous effective mass is not trivial, as it depends not only on the robot’s physical inertial properties but also on its kinematic configuration, the elasticity of joints and links, contact compliance, and, in some cases, the control strategy employed.
ISO/TS 15066 proposes using the quasi-static approximation of the robot’s reflected mass:
\begin{equation}
    \label{eq:iso_effective_mass}
    m_R^\mathrm{ISO} = M_\mathrm{total}/2 + m_l,
\end{equation}
where $M_\mathrm{total}$ is the total moving mass and $m_l$ the effective payload. However, \cite{kirschner2021notion} experimentally demonstrated that this approximation is often overly conservative, reducing robot productivity and, in certain configurations, failing to capture localized dynamic effects that may pose a hazard to humans. 
A more precise dynamic formulation was proposed in \cite{khatib1995inertial}, where the effective mass along a contact direction $\mathbf{u}\in\mathbb{R}^3$ is computed as:
\begin{equation} \label{eq:reflected_mass}
    m_\mathbf{u}(\mathbf{q}) = \left(\mathbf{u}^\top \Lambda_\nu^{-1}(\mathbf{q}) \mathbf{u}\right)^{-1},
\end{equation}
with $\Lambda_\nu^{-1}(\mathbf{q})$ being the upper $3 \times 3$ block of the Cartesian mass matrix inverse:
\begin{equation*}
    \Lambda^{-1}(\mathbf{q}) = J(\mathbf{q}) M^{-1}(\mathbf{q}) J(\mathbf{q})^\top,
\end{equation*}
where $J(\mathbf{q}) \in \mathbb{R}^{3 \times n}$ is the Jacobian at the contact point, $M(\mathbf{q}) \in \mathbb{R}^{n \times n}$ is the joint-space inertia matrix, and $\mathbf{q} \in \mathbb{R}^n$ denotes the joint positions. This formulation explicitly accounts for the robot’s configuration and dynamic coupling, providing a more accurate evaluation of the robot effective mass along directions where collisions may occur.  

In general, the two estimates are not identical, i.e., $m_R \neq m_\mathbf{u}(\mathbf{q})$. The dynamically computed $m_\mathbf{u}(\mathbf{q})$ is typically closer to the real effective mass perceived during an impact, and depending on the configuration, it can be either larger or smaller than the nominal $m_R$. When $m_R > m_\mathbf{u}(\mathbf{q})$, the reduced-mass combination $\mu(\cdot) :=\frac{\cdot + m_H}{\cdot m_H} = \frac{1}{m_R} + \frac{1}{\cdot}$ satisfies $\mu(m_R) > \mu(m_\mathbf{u}(\mathbf{q}))$, which consequently implies that the safe velocity limit computed with the nominal mass is lower,  $v_{0,\max}\big|_{m_R} < v_{0,\max}\big|_{m_\mathbf{u}(\mathbf{q})}$. Conversely, when $m_R < m_\mathbf{u}(\mathbf{q})$, the inverse relation holds. This means that using $m_R$ to define robot motion constraints leads to a more conservative condition (i.e., a lower velocity limit) in configurations where $m_R$ exceeds $m_\mathbf{u}(\mathbf{q})$, and to a potentially unsafe condition (i.e., a higher allowable energy exchange) in configurations where $m_R$ is smaller than $m_\mathbf{u}(\mathbf{q})$.

As discussed in \cite{kirschner2021notion}, \textbf{precise knowledge of the robot's effective mass enables control strategies that enforce safe velocity limits in real time or plan motions that minimize effective mass along likely impact directions, thereby improving safety without unnecessarily constraining robot performance}. For example, \cite{rossi2015pre} demonstrates how kinematic redundancy can be exploited to minimize the reflected mass along the anticipated impact direction, thereby further reducing the risk of injury. Similarly, \cite{steinecker2022mean} introduced the Mean Reflected Mass, a configuration-dependent metric that captures the average reflected mass across all directions, which can be used to evaluate and optimize robot posture for safer operation.
However, the computation relies on accurate models of $M(\mathbf{q})$ and $J(\mathbf{q})$, which may be uncertain due to variable payloads, unmodeled compliance, or friction.


\section{Energy in Safe pHRI} \label{sec:energy}
The core idea underlying the constraints \eqref{eq:constr_deltak}-\eqref{eq:constr_v} is to relate the \emph{maximum elastic energy that can be stored in human tissues} without exceeding tolerable pain thresholds to the \emph{change in the robot’s kinetic energy} between the instant immediately preceding contact and the instant of maximum tissue deformation, as formalized by the energy balance in \eqref{eq:energy_balance_impact}. This relationship leads to operational safety constraints that can be expressed either as an upper bound on the robot velocity \eqref{eq:constr_v} or, equivalently, as a limit on the robot kinetic energy \eqref{eq:constr_k}. As a result, \textbf{mechanical energy emerges as a central quantity} for translating safety requirements into operational constraints on robot motion, supporting both safety assessment and controller implementation. 
This consideration rises to two central research questions: 1) \textit{Is an exclusive focus on mechanical energy sufficient to ensure safety?} and 2) \textit{Do energy-based safety control strategies offer concrete advantages over alternative control approaches?} In the first part of this section, we examine the possibility of assessing safety directly via energy thresholds, bypassing force-/pressure-based measures associated with pain sensitivity. We review studies commonly cited in pHRI literature, highlighting their strengths and limitations compared to traditional force/pressure approaches. In the remainder of the section, we briefly present relevant methods for enforcing energy limits. Rather than attempting a comprehensive survey of all safe pHRI control strategies, which have already been addressed in other works (e.g., \cite{proia2021control,li2024safe}), this discussion focuses specifically on approaches that explicitly regulate  mechanical energy, in particular, the constraints defined in \eqref{eq:constr_deltak}. The aim of this section is to provide exemplary strategies and practical tools for implementing energy limitations, discussing their advantages and inherent limitations, and guiding researchers in identifying open challenges and promising directions for future research.

\subsection{Injury Thresholds} \label{sec:energy_bio_lims}

While the previous sections focused on safety criteria based on force/pressure limits associated with maximum pain tolerance, an alternative line of work characterizes the severity of human safety hazards in terms of the maximum mechanical energy absorbed by the human body up to irreversible damage, shifting the focus from instantaneous peak forces to energy-based exposure. These thresholds were adopted in several early research works preceding the publication of ISO/TS 15066, primarily because a substantial body of experimental evidence on injury biomechanics (originating from automotive safety and impact studies) was already available in energy-based form. In this section, we review these energy-based safety limits, analyze their experimental foundations and modeling assumptions, and critically assess their applicability to physical human-robot interaction, with particular emphasis on their role as global safety constraints rather than direct surrogates for pain perception.

Early experimental studies on cranial and cervical tissues quantified injury tolerance by estimating the energy required to induce structural \textit{failure} (i.e., the point at which the tissue undergoes irreversible damage). For example, skull fracture thresholds were determined by integrating the area under experimentally measured stress-strain curves up to the failure point, referred to as the \textit{energy-to-failure}. Within the tested strain-rate range, this energy-to-failure showed relatively little variation compared to other mechanical parameters such as peak force or ultimate stress \cite{wood1971dynamic}. Similar approaches were adopted for cervical spine components, where the energy to failure under axial tension was approximated as one half of the product of peak force and failure displacement, assuming a linear force-displacement response \cite{yoganandan1996human}. In infant skull specimens, the energy absorbed up to fracture was computed from force-deflection or force-displacement curves and normalized by specimen volume, providing a measure of the tissue’s capacity to dissipate mechanical energy prior to failure \cite{margulies2000infant}.

Beyond purely energy-based metrics, dynamic and power-based injury criteria have also been proposed. The Head Impact Power (HIP) metric estimates the minimum power required to induce injury based on concussion probability in head impacts \cite{newman2000proposed}. Related work suggested that head injury severity scales with the rate of kinetic energy transfer during impact, drawing conceptual parallels with viscous injury criteria in which tissue damage depends on both deformation magnitude and deformation rate \cite{dilorenzo1976power}. 

From a modeling perspective, energy-based thresholds potentially offer several practical advantages. They aggregate the effects of force magnitude and interaction duration into a single scalar quantity, capturing cumulative mechanical load in quasi-static or sustained contacts. In principle, they can reduce the dependence on detailed contact models and tissue-specific stiffness assumptions, which are typically required when translating force or pressure limits into robot motion constraints, as discussed in \autoref{sec:robot_constraints}. Direct energy-based thresholds could also provide a global “budget” for dynamic safety control in robotics, independent of specific contact dynamics such as velocity profile, dwell time, or contact area. These properties make energy-based metrics appealing as high-level safety indicators, particularly in scenarios where contact geometry or tissue compliance is difficult to estimate accurately. However, existing energy- based injury thresholds have important limitations. \textbf{Direct measurements of kinetic energy typically correlate with injury risk rather than isolated pain perception, because pain does not scale linearly with energy}: nociceptor activation depends on strain rate, deformation gradient, and mechanoreceptor density. For instance, the same energy delivered by a needle (highly localized, high pressure $>1000\, \mathrm{N/cm^2}$) triggers sharp A-delta pain via rapid TRPV1/ASIC activation, whereas a foam ball (distributed, low pressure $<50\, \mathrm{N/cm^2}$) dissipates elastically without exceeding neural firing rates, producing dull or no pain. While injury thresholds have the advantage of exhibiting lower inter-subject variability, as they are derived from structural failure tests or objective injury criteria rather than subjective self-assessments, their applicability to mild or pre-injury conditions remains limited. Accurate data on mild injury thresholds are largely unavailable, since existing estimates mainly originate from post-mortem specimens, anthropomorphic test devices, and accident reconstruction studies, which do not reliably capture the onset of non-severe tissue damage in living humans. Moreover, because pain thresholds are lower than mild injury thresholds, they can be adopted as a conservative safety criterion, under the assumption that external pressures not exceeding the pressure pain threshold will not cause any physical injury.


\subsection{ Safe Control Strategies}

Whether safety hazard severity thresholds are defined directly in terms of energy, as in \autoref{sec:energy_bio_lims}, or inferred from force or pressure limits by mapping them to an equivalent potential energy, as in \autoref{sec:robot_constraints}, the resulting methods ultimately constrain the robot’s kinetic energy. This has motivated a substantial body of research on energy-centric safety mechanisms for pHRI. Such mechanisms have the significant benefit of being inherently compatible with \textit{passivity-based control schemes} \cite{van2000l2}, thereby facilitating formal stability proofs, enabling the coordinated design of task and safety objectives, and allowing stable interaction with unknown or uncertain environments.
Within the context of these energy-based control approaches, the notion of \emph{passivity} plays a central role. Passivity is a system-theoretic property that, roughly speaking, ensures that a dynamical system does not autonomously generate an unbounded amount of energy \cite{van2000l2}. This property is stronger than standard notions of stability and is highly desirable for closed-loop control in pHRI for several reasons \cite{stramigioli2015energy}. A particularly relevant aspect for pHRI is that the environment with which the robot interacts cannot, in general, be assumed to be known or accurately modeled during task execution. By enforcing passivity at the robot level, stability of the closed-loop system can be guaranteed for the interconnection with any passive environment, independently of its specific dynamics. Multiple approaches have also moved beyond a purely passivity-based formulation by explicitly enforcing safety through bounds on the energy transfer required to satisfy the constraint in \eqref{eq:constr_k}. Consistent with the modeling abstraction introduced in \autoref{sec:robot_constraints}, these methods rely on the same one-dimensional interaction representation, in which the full dynamics of a multi-articulated robotic system in contact with a human body are mapped onto the direction of contact

\paragraph{Methods overview}

Early energy-based safety strategies in pHRI explored limiting the robot’s stored energy by directly modulating the commanded trajectory, as in \cite{laffranchi2009safe}, which implemented an energy-regulation scheme on a single passive compliant actuator. An inner position controller drives the actuator, while the energy regulation controller monitors the energy content and adjusts the reference trajectory to maintain acceptable energy levels. 
%
%
Subsequent works shifted toward modulating the stiffness of an impedance-controller to keep kinetic energy below predefined thresholds. A notable strategy, originally formulated for a single degree of freedom (DoF) system \cite{tadele2014combining} and subsequently generalized to multi-DoF configurations \cite{raiola2018development}, constrains both the robot’s kinetic energy and the potential energy stored in an impedance controller by appropriately modulating the controller’s stiffness. 
This method has also been extended to series-elastic actuation in \cite{roozi2020energy}, where the authors address the challenges arising from the indirect coupling between the control input and the controlled variable, which is characteristic of elastic actuation.
Building on the same conceptual framework, \cite{geravand2016port} derives appropriate scalings of stiffness to restrict the maximum energy that can be transferred to a human during object co-manipulation. Leveraging the port-Hamiltonian formalism to model power-preserving interconnections among robots, humans, and objects, this approach enables explicit monitoring of energy flows across subsystems, facilitating the implementation of robust energy limitation.
%
%
Other studies propose the integration of energy-based safety indices into optimization frameworks to adaptively modulate robotic motion execution. For example, \cite{rossi2015pre} combines a trajectory generator with a constraint-based optimization algorithm that feeds a lower-level position/velocity controller. The general idea of this approach is to ensure that the energy potentially dissipated during collisions remains below a prescribed threshold while the robot follows a given path. The optimization problem considers all combinations of potential collision points between different robot links and human body regions, subject to multiple constraints. In addition, kinematic redundancy is exploited to minimize the reflected mass along anticipated impact directions, further reducing the risk of injury.
Similarly, \cite{meguenani2015control} incorporates transferred energy as a quadratic constraint in an optimization-based controller, adaptively scaling trajectory execution to satisfy energy limits and modulating the constraint according to the human-robot separation distance.  
This approach is extended in \cite{meguenani2016energy}, where a complementary safety indicator based on stored potential energy is introduced to capture post-impact contact forces. Both kinetic and potential energy indicators are integrated into a quadratic optimization framework for torque control, enabling trajectory tracking under energetic safety constraints.

%
%
More recent works have instead proposed the use of virtual energy tanks between the controller and the robot. 
A \emph{virtual energy tank} acts as an artificial energy storage element that monitors and regulates the energy flow between the controller and the physical system and is typically employed to ensure passivity  
 on the closed-loop system \cite{califano2022use}. 
 In the standard formulation of a virtual energy tank (i.e., without energy-recycling mechanisms), the tank’s initial energy specifies the upper bound on the amount of energy that the controller can deliver to the robot and exchange with the environment. Consequently, to satisfy safety requirements, the tank is initialized with an energy level below the maximum energy that could be transferred to a human during an impact, thereby directly defining the upper bound on the admissible energy throughout task execution.
 Leveraging this concept, the approach presented in \cite{michel2022safety} enforces kinetic-energy-based safety constraints within a hierarchical robot controller by coupling it with a virtual energy tank that explicitly bounds the energy available to the control actions. The framework is formulated for time-varying compliance control in the regulation case and is generalized to accommodate an arbitrary number of prioritized tasks while preserving the passivity of the overall closed-loop system through the use of energy tanks. In this setting, kinetic energy constraints are enforced via the tank: the initial energy level is selected to limit the maximum energy that can be injected into the robot during free motion. Since the framework also accounts for refilling the tank through dissipative effects, the tank cannot be simply initialized at the prescribed safe energy limit. Instead, they need to estimate the nominal kinetic energy of the robot during motion. Consequently, the initial tank energy is set equal to the safe energy threshold minus the total elastic potential energy associated with all constant-stiffness elements included in the controller.
 Similarly, in \cite{benzi2023energy} energetic safety constraints are enforced via a convex optimization problem that filters nominal control inputs according to the remaining energy in a virtual tank, with a discrete-time implementation suitable for real-time use. The authors further introduce augmented tank dynamics that explicitly account for external energy exchanges while preserving convexity: external energy injections (e.g., human pushes) are routed into the tank to prevent safety violations, whereas unmodeled energy extractions (e.g., contact-induced decelerations) are handled to avoid unnecessary performance loss. The framework also allows switching energy bounds, for example based on the closest human body region, enabling context-aware safety regulation.
Using energy tanks, energy bounds can also be easily enforced in learning-based controllers. For instance, \cite{zanella2024learning} demonstrates that a deep reinforcement learning agent can learn to remain under a desired energy threshold, thereby implicitly preserving passivity of the closed-loop system while accomplishing the task. Since the energy limit is explicitly set by the tank initialization, this approach can also ensure that the agent respects safe energy-level constraints during both learning and execution.
 
%
%
A class of tools that has recently seen wide adoption  due to its strong theoretical guarantees and algorithmic simplicity is represented by \emph{Control Barrier Functions} (CBFs) \cite{Ames2019ControlApplications, Ferraguti2022SafetyApproach}. CBFs provide a formal mechanism to constrain the evolution of a controlled system so that its state remains within a prescribed safe set. In many robotic applications, safety specifications inherently conflict with task execution objectives. This leads to a semantic separation between the main task, encoded in an objective function, and safety specifications, expressed as constraint. Such a separation has motivated the notion of \emph{safety filters}  \cite{Wabersich2023Data-DrivenSystems}. A safety filter is a computational layer that takes as input a nominal control action, designed to achieve the primary task while neglecting safety, and returns a modified control input that (i) minimally deviates from the nominal command and (ii) guarantees satisfaction of the safety constraints. CBFs are among the most popular implementations of safety filters. In the context of safe pHRI, CBFs are primarily employed to enforce the velocity limit \eqref{eq:constr_v}, which is equivalent to the kinetic energy limit \eqref{eq:constr_k}, as previously discussed in this document. 
 For example, a CBF-based control strategy has been proposed in \cite{cortez2021safe} to ensure passivity while enforcing velocity constraints, thereby ensuring safe pHRI in compliance with the standard regulations. This method systematically incorporates the robot’s energy dynamics into the barrier function design, adapting the \emph{energy-based CBFs} framework proposed in \cite{singletary2021safety} to resolve potential ill-posedness that may arise in pHRI, consequently ensuring the safety, passivity, and robustness of mechanical systems operating in the presence of humans. 
 Nevertheless, CBFs also provide an effective mechanism for directly constraining a system’s energy, thereby offering formal guarantees on admissible energy bounds, as demonstrated in \cite{califano2024effect}. In that work, energy-based CBFs are employed to impose constraints on both the kinetic and total energy of a generic mechanical system, together with a systematic methodology for assessing the influence of safety-critical control schemes on the system’s overall energy balance. 

%
%
Recent research highlights that limiting a robot’s total energy alone is not sufficient to reduce the severity of impacts with humans. In fact, even when energy bounds are respected, excessive power during contact can produce sudden, hazardous motions. Consequently, constraining the rate at which energy is delivered, i.e., the robot’s power output, is equally crucial for safety. This insight has motivated the development of a variety of control strategies that explicitly regulate power during physical interaction.
One solution that has been explored leverages the controller parameters of impedance controllers, in a manner similar to approaches used for energy limitation. In fact, the same works that proposed limiting energy through stiffness also suggest limiting power via damping \cite{tadele2014combining, raiola2018development, geravand2016port}. This approach has been further extended to continuous pHRI using a port-Hamiltonian framework \cite{geravand2016port}, where power ports explicitly enforce the maximum allowable power.
Another line of research proposes extending virtual energy tank frameworks to meet power-related control objectives. Central to this approach is the concept of power flow regulation, implemented via power valves that limit the extractable power from the ports of the energy tank \cite{shahriari2018valve}. 
As demonstrated in \cite{shahriari2019power}, regulating the rate of energy exchange to follow a prescribed reference power trajectory not only facilitates smooth and safe power transfer but also mitigates the challenge of appropriately sizing the energy tank. Insufficient initial energy may cause the passivity-violating control component to disengage before the task is completed, whereas an excessively large energy allocation may permit the controller to deviate from the equilibrium point or enable certain subsystems to accumulate undesirably high energy levels. Appropriately designed power regulation thus guarantees that energy is available when required, without compromising system stability or operational safety.
The concept of valve-based virtual energy tanks has been further extended in \cite{michel2022safety} to redundant manipulators controlled by variable-impedance controllers, where real-time power injection is distributed across tasks via convex optimization. This allows higher-priority tasks to maintain stability under energy constraints while respecting task-specific power limits.
In \cite{michel2024novel}, higher-order CBFs are integrated into a novel virtual energy tank formulation that models the tank with second-order dynamics rather than a simple integrator. This formulation avoids the discrete switching behavior typical of classical energy tank methods, which can cause control chattering. The CBFs provide formal guarantees for enforcing prescribed power constraints, enabling the computation of an optimal control input that maintains the system state within a safe set in which both tank energy and power are constrained. Although this approach permits the specification of an upper bound on power (for safety) and a lower bound on energy (to ensure task completion), it does not allow simultaneous strict enforcement of explicit upper bounds on both energy and power; at any given time, only one of these quantities can be tightly constrained. While the energy remains bounded by virtue of passivity, the corresponding upper bound cannot be explicitly set, which complicates the verification of compliance with ISO/TS 15066.
Since these approaches rely on fixed maximum power bounds, which can be overly conservative in unstructured environments, \cite{benzi2023unified} propose an iterative learning control scheme that simultaneously adapts the admittance dynamics and the exertable power bounds based on the directional tracking error. Here, power-based saturation is embedded directly in the adaptive control laws, preventing aggressive behavior while ensuring compliant and effective pHRI.

%
%
%

\paragraph{Remark on Energy Tanks, Safety and Passivity}
 
The energy tank framework offers a systematic methodology for enforcing passivity in arbitrary control actions, independently of the system’s dynamics \cite{califano2022use}. Passivity is formally guaranteed as long as the tank is initialized with finite energy. However, in practice, accurately estimating the appropriate initial energy can be challenging \cite{zanella2024learning}, since the required level depends on the specific task and the control actions executed. Although initializing the tank with a very large (yet bounded) amount of energy formally preserves passivity, such a choice can induce so-called “practically unstable” behaviors \cite{benzi2022energy}.
As a consequence, a naive tank design \textit{de facto} renders robust stability a property that is decoupled from any explicit safety guaranty. Specifically, as long as the energy stored in the tank is not fully depleted, the control input remains completely transparent to the tank algorithm. In this situation, the tank mechanism essentially degenerates into a trick to formally prove passivity, with only limited practical relevance in contexts where tasks must be executed in unstructured environments.
This observation is also associated with the frequently encountered misconception that passivity is equivalent to safety \cite{folkertsma2018safety, music2018passive,papageorgiou2020passive}. Passivity is a desirable system property because it is required to guaranty stable interactions with arbitrarily unknown environments. However, passivity alone does not ensure safe operation, since a system with high (yet bounded) stored energy can still produce dangerous power ejections over short time intervals. 

\paragraph{Remark on Energy Constraint}

The existing body of literature often lacks clarity regarding the precise formulation of energy-based safety constraints. Several works published before the release of ISO/TS 15066 adopt explicit injury-related energy thresholds. In \autoref{sec:energy_bio_lims}, we have already discussed the potential advantages and the inherent challenges associated with the use of these values. The majority of works published after the technical specification refer more generically to limiting the robot’s kinetic energy, often without explicitly stating the associated threshold. Some work imposes that the robot’s kinetic energy must not exceed the maximum elastic potential energy that a human body region can safely absorb, formally constraining $K_0 \leq U_{s,\max}$ \cite{benzi2023energy}. This constraint differs from what we derived in \autoref{sec:robot_constraints} and reflects a distinct and more conservative modeling assumption, as it implicitly assumes an infinitely large human effective mass, equivalent to a \textit{potentially-clamped scenario} in which no body motion occurs, as discussed in \autoref{sec:human_effective_mass}. This represents an extreme case that, while possible, is far from general and should be treated as a separate condition, as explicitly foreseen by the technical specification. In reality, the human body can absorb only a portion of the robot’s energy depending on its effective mass and compliance. As a result, this assumption systematically overestimates the risk, leading to unnecessarily conservative safety limits and penalizing robot performance. 
%


\section{Final Discussion}

\begin{figure*}
    \centering
    \includegraphics[width=0.99\linewidth]{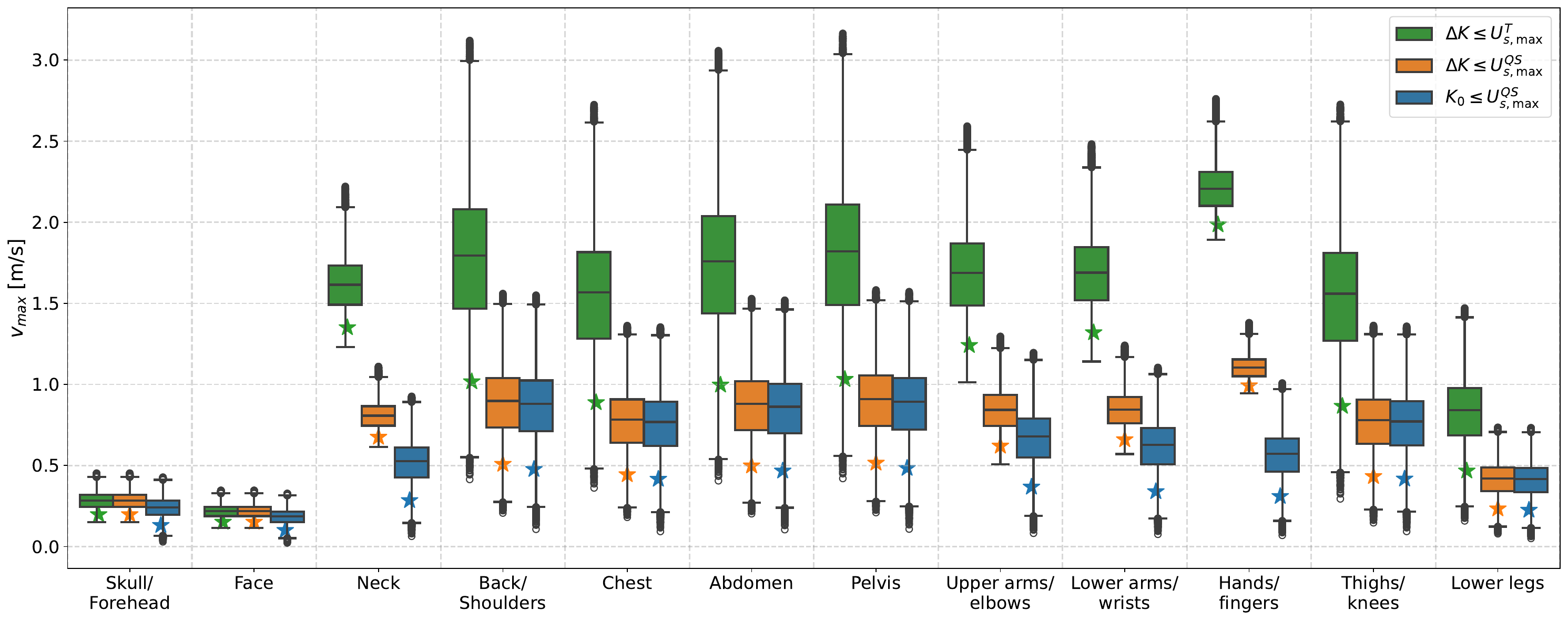}
    \caption{Admissible end-effector velocities across ISO body regions for three safety formulations: transient  (T) contact with scaled force/pressure limits (green), quasi-static (QS) contact following ISO/TS 15066 limits (orange), and the potentially-clamped scenario using \eqref{eq:clamped_scenario} (blue). Boxplots show the variability of velocity limits resulting from the configuration- and direction-dependent apparent mass $m_\mathbf{u}(q)$, while star markers ($\bigstar$) indicate the corresponding scalar limits obtained using the constant ISO-based mass $m_R^{ISO}$.}
    \label{fig:boxplots}
\end{figure*}

It has been extensively emphasized in the document that the safety constraints on robot motion depend critically on knowledge of the human body part that may be involved in a contact and on whether the interaction happens with a human who can recoil or not, as discussed in \autoref{sec:human_effective_mass}. Different regions of the body exhibit distinct properties, such as tissue stiffness, effective mass, and pain sensitive thresholds, all of which affect the maximum safe robot speed/energy. Similarly, the potentially-clamped scenario condition imposes stricter limits because the human segment is immobilized and therefore cannot attenuate via motion the transfer of energy into tissue deformation, requiring lower permissible velocities and kinetic energy.
Ideally, a robotic system should be equipped with a perception system capable of reliably monitoring both environmental dynamics and human motion, enabling dynamic adaptation of speed/energy limits to the context. In particular, if the system can identify the specific human body region involved in a potential collision, limits can be adjusted accordingly. For instance, a collision with the chest allows higher permissible speed/energy compared to a collision with the face, where stricter limits are required. Similarly, if the environment is free of obstacles that could create a clamped scenario, the robot may move faster, using constraint in \eqref{eq:constr_v} or \eqref{eq:constr_k} with transient force/pressure limits. Conversely, when the possibility of a clamped contact is detected, safety limits should be reduced according to \eqref{eq:clamped_scenario}, with more conservative quasi-static force/pressure thresholds.
Nevertheless, most industrial setups, do not implement such advanced perception-and-adaptation systems. This is partly due to the limited maturity and reliability of current technology, and partly because many SMEs prefer simpler, cost-effective solutions. In these cases, safety is ensured by assuming worst-case scenarios: if a clamped scenario is possible, speed and energy limits are set according to \eqref{eq:clamped_scenario}; if no human tracking is available, limits correspond to the most conservative body region, typically the face. While this approach significantly reduces performance, it guarantees compliance with safety thresholds and provides robust protection under all possible interactions.

Building on the discussion developed so far, we can identify certain key design choices that directly affect safety and performance in pHRI trading off system complexity. The first choice is technological: whether to integrate a perception-based, context-aware strategy that can dynamically adapt speed and energy limits according to the human operator’s position and the surrounding environment, or to keep the setup simple and rely on conservative, worst-case assumptions. The former approach can maximize performance by allowing the robot to move faster when conditions are safe, while the latter prioritizes reliability and ease of implementation by always assuming the most restrictive scenario.
Another choice concerns the robot’s effective mass. As discussed in \autoref{sec:robot_effective_mass}, using the configuration-dependent apparent mass $m_\mathbf{u}(q)$ defined in \eqref{eq:reflected_mass} provides a more precise estimate along the contact direction, improving both performance and safety compared to the constant approximation $m_R^{ISO}$ proposed in ISO/TS 15066 (\eqref{eq:iso_effective_mass}). However, the use of $m_\mathbf{u}(q)$ introduces additional computational complexity, which may need to be managed in real-time applications. In particular, it depends on the joint configuration $q$,  the inertia matrix $M(q)$, and a perception system able to reliably estimate the contact direction $\mathbf{u}$.

To conclude the discussion, we study the effects of the design choices outlined above by analyzing how the pre-collision safe speed limits of the robot change under different assumptions. While more conservative limits are expected in potentially-clamped scenarios and when considering collisions with highly sensitive body parts, such as the face, we aim to provide numerical references to quantify the potential performance loss compared to an advanced system capable of dynamically adapting limits to the context.
For this analysis, we consider a Franka Emika Panda manipulator, with inertial parameters from \cite{gaz2019dynamic}, and no end-effector or payload is included. Using the quasi-static approximation proposed in ISO/TS 15066, the resulting reflected mass is $m_R^{ISO} = 5.545,\mathrm{kg}$. We also compare speed limits for each body part using the configuration-dependent apparent mass $m_\mathbf{u}(q)$, where admissible limits vary with the robot configuration.
To evaluate this, the workspace is discretized on a uniform 
\(5\,\text{cm}\)  grid, using a single end-effector orientation with the flange pointing downward and the frame aligned with the world axes. For each grid position, a joint configuration is computed via inverse kinematics \cite{corke2017robotics}, and the reflected mass is evaluated along 20 uniformly distributed Cartesian directions $u$. 

In \autoref{fig:boxplots}, we report the pre-collision speed limits across ISO-defined body regions under different safety constraint options. In particular, we compare quasi-static contact conditions, which include potentially-clamped scenarios, with transient contact conditions (in which clamping is excluded).
For the quasi-static contact, two alternative constraints are considered. The first follows the approach suggested in ISO/TS 15066, which applies the kinetic energy constraint in \eqref{eq:constr_k} without explicit assumptions on human mobility,  i.e., $\Delta K \leq U_{s,\max}^{QS}$, where $U_{s,\max}^{QS}$ denotes the maximum admissible elastic energy derived from quasi-static thresholds. The second adopts the model in \eqref{eq:clamped_scenario}, which explicitly assumes an immobilized human body part. In this case, the entire pre-impact kinetic energy may be transferred to the human, leading to the stricter condition $K_0 \leq U_{s,\max}^{QS}$.
For the transient contact, we consider the standard constraint \eqref{eq:constr_k} as specified in ISO/TS 15066, i.e., $\Delta K \leq U_{s,\max}^{T}$, where $U_{s,\max}^{T}$ is derived from force and pressure limits scaled with respect to quasi-static values. In accordance with the technical specification, these transient limits are obtained by applying a multiplicative factor of two to the quasi-static values for all body regions, with the exception of the skull/forehead and the face, for which no scaling is applied.
Within the same figure, we also compare speed limits obtained using two different estimates of the robot’s effective mass: the constant value $m_R^{ISO}$ and the configuration-dependent apparent mass $m_\mathbf{u}(q)$. While $m_R^{ISO}$ yields a constant speed limit for each body region, $m_\mathbf{u}(q)$ varies with the robot's configuration and contact direction. Consequently, the latter is visualized using boxplots, which summarize the spread across the sampled configurations.

As expected, limits related to transient contact are on average higher than the other formulations across all body regions, reflecting the relaxed constraints when clamping is not expected. In potentially-clamped scenarios, the mobility-dependent constraint $K_0 \leq U_{s,\max}^{QS}$ yields lower average speed limits than the ISO-recommended quasi-static formulation $\Delta K \leq U_{s,\max}^{QS}$, indicating that the ISO approach may underestimate the risk in these situations. The largest discrepancies between these two constraints occur for the neck, hands/fingers, and arms, regions that combine lower tissue stiffness with smaller effective mass, making them more sensitive to the assumption of human immobility. Finally, using the constant mass $m_R^{ISO}$ produces generally lower speed limits for each body region compared to the configuration- and direction-dependent apparent mass $m_\mathbf{u}(q)$, highlighting the performance gains achievable when accounting for the robot’s dynamic inertia.

\begin{table}[t]
\centering 
\caption{Pre-collision average speed limits $v_\mathrm{max}^{avg}$ and relative percentage scaling for different body regions, derived under design simplifications on the robot’s effective mass $m_R$ and contact conditions (transient or quasi-static). }
\label{tab:v0max_scaling}
\begin{tabular}{lcccc} 
 & $v_\mathrm{max}^{avg}\, [\mathrm{m/s}]$ & \multicolumn{3}{c}{Speed Scaling [\%]} \\ 
\cmidrule(lr){2-5}
  \textbf{Body Region} & $ \begin{matrix}
\Delta K \leq U_{s,\max}^{T} \\ m_R=m_\mathbf{u}(q)
\end{matrix}$ & $ \begin{matrix}
\Delta K \leq U_{s,\max}^{T} \\ m_R=m_R^{ISO}
\end{matrix}$  & $ \begin{matrix}
 K_0 \leq U_{s,\max}^{QS} \\ m_R=m_\mathbf{u}(q)
\end{matrix}$  & $ \begin{matrix}
K_0 \leq U_{s,\max}^{QS} \\ m_R=m_R^{ISO}
\end{matrix}$  \\ 
\midrule
Skull/Forehead & 0.283 & 70 & 85 & 47 \\
Face & 0.217 & 70 & 85 & 47 \\
Neck & 1.615 & 84 & 32 & 18 \\
Back/Shoulders & 1.770 & 57 & 49 & 27 \\
Chest & 1.546 & 57 & 49 & 27 \\
Abdomen & 1.735 & 57 & 49 & 27 \\
Pelvis & 1.795 & 58 & 49 & 27 \\
Upper arms/elbows & 1.681 & 74 & 40 & 22 \\
Lower arms/wrists & 1.687 & 78 & 37 & 20 \\
Hands/fingers & 2.209 & 90 & \textbf{26} & \textbf{14} \\
Thighs/knees & 1.536 & \textbf{56} & 49 & 27 \\
Lower legs & 0.828 & \textbf{56} & 49 & 27 \\
\bottomrule
\end{tabular}
\end{table}

\begin{table}[t]
\centering
\caption{Percentage reduction in admissible pre-collision speed limits relative to the average transient-contact reference case $\Delta K \leq U_{s,\max}^{T}$ computed with the configuration-dependent effective mass $m_R = m_\mathbf{u}(q)$, when worst-case assumptions on the contacted body region are enforced..}
\label{tab:speed_reduction_contact_type}
\begin{tabular}{lcc}
& \multicolumn{2}{c}{Worst-case Speed Scaling [\%]} \\ 
\cmidrule(lr){2-3}
\textbf{Body region} &   $ \begin{matrix}
\Delta K \leq U_{s,\max}^{T} \\ m_R=m_R^{ISO}
\end{matrix}$ & $ \begin{matrix}
K_0 \leq U_{s,\max}^{QS} \\ m_R=m_R^{ISO}
\end{matrix}$  \\
\midrule
Skull/Forehead & 54 & 35 \\
Face & 70 & 46 \\
Neck & \textbf{9} &\textbf{6} \\
Back/Shoulders & \textbf{9} & \textbf{6} \\
Chest & \textbf{10} & \textbf{6} \\
Abdomen & \textbf{9} & \textbf{6} \\
Pelvis & \textbf{8} & \textbf{6} \\
Upper arms/elbows & \textbf{9} & \textbf{6} \\
Lower arms/wrists & \textbf{9} & \textbf{6} \\
Hands/fingers & \textbf{7} & \textbf{5} \\
Thighs/knees & \textbf{10} & \textbf{7} \\
Lower legs & 18 & 12 \\
\bottomrule
\end{tabular}
\end{table}

When human motion and environmental context cannot be reliably tracked, the robot must operate under fixed and conservative safety limits, without knowledge of the contact direction, the involved body region, or the possibility of clamping conditions. To quantify the conservatism introduced by these uncertainties, we define baseline reference values as the average admissible pre-collision speed limits $v_\mathrm{max}^{avg}$ under the transient contact assumption (i.e., no clamping expected), computed using the configuration- and direction-dependent effective mass of the robot $m_R = m_{\mathbf{u}}(q)$. These values correspond to the least conservative case and are reported in the first column of  \autoref{tab:v0max_scaling}. The remaining columns of the  table quantify how these baseline limits are reduced when introducing progressively more conservative conditions. Specifically, we consider two sources of conservatism: (i) replacing the configuration- and direction-dependent effective mass with the constant approximation from ISO/TS 15066 ($m_R = m_R^{ISO}$), and (ii) additionally accounting for a persistent possibility of clamped contact through the constraint $K_0 \leq U_{s,\max}^{QS}$. The second column reports the effect of assumption (i) alone, the third column reports the effect of assumption (ii) alone, and the forth column reports the combined effect of both conditions. The resulting \textit{Speed Scaling} [\%] values therefore quantify the reduction in admissible pre-collision speed induced by each modeling simplification relative to the baseline physically informed case. 
The largest reductions occur for peripheral and highly mobile segments, such as the hands, fingers, and arms, where \textbf{admissible speeds can fall below $\mathbf{15 \%}$} of the baseline, whereas larger, less mobile regions like the torso and pelvis are comparatively less affected. 
When the controller has no information about which human body region may be involved in a potential collision, an additional loss in admissible speed arises. In this case, safety constraints cannot be adapted to the interaction context and must be enforced with respect to the most restrictive anatomical region, namely the face. This worst-case assignment propagates to all possible contacts, practically replacing region-specific limits with face-based constraints. 
Under the ISO effective mass approximation $m_R = m_R^{ISO}$, the baseline admissible speed for a safe contact with the face (from \autoref{tab:v0max_scaling}) is approximately $0.217\  \mathrm{m/s}$. Using the speed scaling values reported in the second and fourth columns of \autoref{tab:v0max_scaling}, the corresponding worst-case admissible speeds are obtained as scaled versions of this baseline: $0.15\ \mathrm{m/s} = 0.70 \cdot 0.217\  \mathrm{m/s}$ for transient contact and $0.10\  \mathrm{m/s} = 0.47 \cdot 0.217\  \mathrm{m/s}$ for quasi-static contact.
\autoref{tab:speed_reduction_contact_type} reports the resulting \textit{Worst-case Speed Scaling} [\%] for each body region under the same worst-case assumption, i.e., assuming potential collision with the face relative to the baseline values $v_\mathrm{max}^{avg}$. This quantifies the additional degradation introduced when body-part information is unavailable and all contacts must be treated as face-equivalent.
This gap highlights the practical importance of context-aware safety evaluation. With configuration-dependent effective mass, correct identification of low-sensitivity contact regions (e.g., arm, torso, or abdomen), and under the assumption of non-clamped interaction, the robot\textbf{ can operate more than $\mathbf{10\times}$ faster} while remaining within safety limits compared to a conservative setting in which a constant effective mass approximation is used and every potential collision is treated as involving the most sensitive region (the face), together with a persistent possibility of clamped contact.

\section{Conclusion and Future Research Directions} \label{sec:conclusion}

Cobots are reshaping contemporary manufacturing by allowing humans and robots to share workspaces in a safe and efficient manner. In contrast to conventional industrial robots that operate behind protective barriers, cobots are intended for open settings where direct contact with humans can occur. Ensuring safety in pHRI constitutes a significant challenge, as it is affected not only by the robot’s control and motion but also by human factors, task-specific requirements, and surrounding environmental conditions. ISO 10218-2 and ISO/TS 15066 specify quantifiable safety constraints for robot motion, but these values are highly context-sensitive and must be interpreted cautiously to prevent underestimation of risks and to mitigate potential degradation in robotic system performance.

In this paper, we provide a systematic derivation of the safety constraints for the PFL mode defined in ISO/TS 15066. All underlying assumptions were explicitly stated and motivated, and the derivation process was analyzed in detail to clarify how robot parameters, human body-region characteristics, and interaction conditions determine admissible pre-collision velocities and energies. This formalization provides a rigorous foundation for assessing pHRI safety and understanding the operational meaning of the standard’s recommendations. \textbf{Future work} should move beyond simplified lumped-mass models, incorporating detailed representations of the human musculoskeletal chain and the robot kinematic chain. Such models would capture multi-contact dynamics, complex energy transfer, and interactions that cannot be represented by a single effective mass and spring.

From this derivation, several key remarks emerged. First, the instantaneous safety condition at maximum tissue deformation is independent of impact restitution. Second, the admissible robot motion critically depends on the pain thresholds of the contacted body region. Beyond tissue stiffness and maximum tolerable force, the effective contact area is central: allowable speed scales linearly with area. \textbf{Current literature lacks } controllers that actively manage contact area during interaction. Smaller contact areas increase local pressure and reduce safe velocities, whereas larger areas permit faster motion without exceeding biomechanical limits. This consideration applies not only to the robot’s structure but also to any attached tools or manipulated objects, complementing region-specific dynamic thresholds and improving upon the conservative ISO/TS 15066 recommendations.  

Third, the maximum safe impact velocity $v_{0,\max}$ depends on the reduced-mass combination of the robot $m_R$ and the human $m_H$. Lower reflected inertia, achieved through lighter robots or favorable configurations, allows for higher admissible velocities. In extreme cases, when $m_R \gg m_H$, the bound is dominated by the human mass, whereas when $m_H \gg m_R$, it is primarily determined by the robot mass and tissue stiffness. Additionally, in the clamped case ($m_H \to \infty$), the absence of human motion prevents energy attenuation, resulting in the most restrictive limits.  Accurate estimation of both human and robot effective masses is therefore essential for realistic safety assessment. In this context, while ISO/TS 15066 adopts conservative values of $m_H$ for transient contact, the constraints prescribed for quasi-static contact with immobilized body segments may not ensure adequate safety margins, and the use of quasi-static robot mass approximations can further lead to velocity limits that are either overly conservative or insufficiently protective.  

Mechanical energy lies at the heart of safe pHRI and serves as a key metric for safety. Unlike subjective force or pressure limits based on pain, energy-based thresholds quantify the mechanical energy leading to injury, with lower inter-subject variability. However, focusing on injury rather than pain makes them less conservative in mild or pre-injury conditions, potentially increasing hazards. The fundamental role of energy has  driven extensive research on energy-centric safety mechanisms. These approaches are inherently compatible with passivity-based schemes, enabling formal stability proofs, coordinated design of task and safety objectives, and stable interaction with unknown or uncertain environments.  However, the existing literature still offers \textbf{opportunities  for further research}. For instance, current solutions that combine energy and power limiting exhibit intrinsic limitations, while their integration with post-impact policies, compliant controllers, or obstacle avoidance strategies remains unexplored. Further research could focus on coupling energy-based safety controllers with AI-driven decision-making to provide formal safety and stability guaranties. In fact, present approaches are often highly task-specific, which complicates integration with AI controllers.

Key design choices affecting safety and performance were identified as a synthesis of the overall analysis and then quantitatively evaluated through numerical examples. The results highlight the central role of reliable perception in the safety-performance trade-off, particularly for tracking human motion and anticipating potentially-clamped scenarios. In fact, safety architectures based on fixed conservative limits (common in PFL implementations) lead to significant performance losses. \textbf{Future systems} should tightly couple real-time perception with energy-based safety controllers, enabling adaptive limits that dynamically respond to task conditions, human motion, and environmental variability. This approach reduces overly conservative constraints while maintaining safety.

Finally, safety for sustained-contact scenarios has received limited attention in current standards. \textbf{Methods are needed} to evaluate safety during continuous physical contact, as occurs in cooperative tasks such as co-assembly, co-transport, or physical guidance. In these cases, risk cannot be captured solely by collision-based metrics, since prolonged contact can produce sustained or repetitive loading on tissues, joints, and muscles, potentially causing strain, discomfort, or overuse injuries that are not addressed by existing standards.

In summary, while significant progress has been made in deriving and implementing safety constraints for pHRI, several open challenges remain. These include improving human-robot interaction models, optimizing contact areas, developing context-aware and adaptive safety architectures, integrating energy-based limits with AI-driven control and obstacle avoidance, and addressing sustained-contact scenarios that cannot be reduced to collision metrics.


\printbibliography 

\end{document}